\documentclass[twocolumn,floatfix,showpacs,prd,aps,tightenlines]{revtex4}
\usepackage{graphicx}
\usepackage{amsmath}
\usepackage{psfrag}
\usepackage{dcolumn}
\usepackage{bm}

\newcommand{\bea}{\begin{eqnarray}}
\newcommand{\eea}{\end{eqnarray}}
\newcommand{\beq}{\begin{equation}}
\newcommand{\eeq}{\end{equation}}
\newcommand{\KMS}{\rm km\,s^{-1}}

\usepackage{color}

\begin{document}

\title{Modeling maximum astrophysical gravitational recoil velocities}

\author{Carlos O. Lousto}
\affiliation{Center for Computational Relativity and Gravitation,
School of Mathematical Sciences,
Rochester Institute of Technology, 78 Lomb Memorial Drive, Rochester,
 New York 14623}

\author{Yosef Zlochower} 
\affiliation{Center for Computational Relativity and Gravitation,
School of Mathematical Sciences,
Rochester Institute of Technology, 78 Lomb Memorial Drive, Rochester,
 New York 14623}

\date{\today}

\begin{abstract}
We measure the recoil velocity as a function of spin for equal-mass,
highly-spinning black-hole binaries, with spins in the orbital plane,
equal in magnitude and opposite in direction. We confirm that the
leading-order effect is linear in the spin and the cosine of angle
between the spin direction and the infall direction at merger. We find
higher-order corrections that are proportional to the odd powers in both
the spin and cosine of this angle. Taking these corrections into
account, we predict that the maximum recoil will be $3680\pm130 \KMS$.
\end{abstract}

 \maketitle

\section{Introduction}

The field of Numerical Relativity (NR) has progressed at a remarkable
pace since the breakthroughs of 2005~\cite{Pretorius:2005gq,
Campanelli:2005dd, Baker:2005vv} with the first successful fully
non-linear dynamical numerical simulation of the inspiral, merger, and
ringdown of an orbiting black-hole binary (BHB) system.
BHB physics has rapidly matured into a critical tool for
gravitational wave (GW) data analysis and astrophysics.  Recent
developments include: studies of the orbital dynamics of spinning
BHBs~\cite{Campanelli:2006uy, Campanelli:2006fg, Campanelli:2006fy,
Herrmann:2007ac,
Herrmann:2007ex, Marronetti:2007ya, Marronetti:2007wz, Berti:2007fi},
calculations of recoil velocities from the merger of unequal mass
BHBs~\cite{Herrmann:2006ks, Baker:2006vn, Gonzalez:2006md}, and
very large recoils acquired by the remnant of the merger of two spinning BHs
~\cite{Campanelli:2007ew, Campanelli:2007cga, Lousto:2008dn, Pollney:2007ss,
Gonzalez:2007hi, Brugmann:2007zj, Choi:2007eu, Baker:2007gi,
Schnittman:2007ij, Baker:2008md, Healy:2008js, Herrmann:2007zz,
Herrmann:2007ex, Tichy:2007hk, Koppitz:2007ev, Miller:2008en},
empirical models relating the final mass and spin of
the remnant with the spins of the individual BHs
~\cite{Boyle:2007sz, Boyle:2007ru, Buonanno:2007sv, Tichy:2008du,
Kesden:2008ga, Barausse:2009uz, Rezzolla:2008sd, Lousto:2009mf}, and
comparisons of waveforms and orbital dynamics of
BHB inspirals with post-Newtonian (PN)
predictions~\cite{Buonanno:2006ui, Baker:2006ha, Pan:2007nw,
Buonanno:2007pf, Hannam:2007ik, Hannam:2007wf, Gopakumar:2007vh,
Hinder:2008kv}.

The surprising discovery ~\cite{Campanelli:2007ew, Campanelli:2007cga}
that the merger of binary black holes can produce recoil
velocities up to $4000\ \KMS$, and hence allow the remnant to escape
from major galaxies, 
led to numerous theoretical and observational
efforts to find traces of this phenomenon. Several studies made
predictions of specific observational features of recoiling
supermassive black holes in the cores of galaxies in the
electromagnetic spectrum \citep{Haiman:2008zy, Shields:2008va,
Lippai:2008fx, Shields:2007ca, Komossa:2008ye, Bonning:2007vt,
Loeb:2007wz} from infrared \citep{Schnittman:2008ez} to X-rays
\citep{Devecchi:2008qy, Fujita:2008ka, Fujita:2008yi} and
morphological aspects of the galaxy cores \citep{Komossa:2008as,
Merritt:2008kg, Volonteri:2008gj}.  Notably, there began to appear
observations indicating the possibility of detection of such effects
\citep{Komossa:2008qd, Strateva:2008wt,Shields:2009jf}, and although
alternative explanations are possible \citep{Heckman:2008en,
Shields:2008kn, Bogdanovic:2008uz, Dotti:2008yb},  there is still the
exciting possibility that these observations can lead to the first
confirmation of a prediction of General Relativity in the
highly-dynamical, strong-field regime.

Numerical simulations of the BHB problem have sampled the parameter 
space of the binary for different values of the binary's mass ratio
$q$ and arbitrary orientations of the individual spins of the holes.
Two astrophysically important regions of this parameter space remain
challenging to describe accurately by numerical simulations: the small
$q$ limit, although recent development of the numerical techniques have
produced a successful simulation of the last few orbits before the merger of
a $q=1/100$ binary~\cite{Lousto:2010ut}, and the near maximal
spin limit.
The most recent simulations of highly-spinning BHBs was
of non-precessing binaries with intrinsic
spins $\alpha=0.95$ \cite{Lovelace:2010ne}.
 Since BHBs with $\alpha=1$ are still elusive to full numerical
simulations, and the configuration that maximizes the gravitational
recoil is one that starts with maximally spinning BHs, with opposite
spins lying on the orbital plane
\cite{Campanelli:2007ew,Campanelli:2007cga}, we will model these
configurations for different values of the intrinsic spin parameter up
to $\alpha=0.92$ (which is achievable with current techniques to solve
initial ``puncture'' data) and then extrapolate to
$\alpha=1$ using an improves version of our original
empirical formula~\cite{Campanelli:2007ew, Campanelli:2007cga,
Lousto:2009mf}.

In Ref.~\cite{Lousto:2009mf} we extended our original empirical formula
for the recoil velocity imparted to the remnant of a 
BHB merger~\cite{Campanelli:2007ew, Campanelli:2007cga} to include
next-to-leading-order corrections, still linear in the spins.
The extended formula has the form:
\begin{eqnarray}\label{eq:Pempirical}
\vec{V}_{\rm recoil}(q,\vec\alpha)&=&v_m\,\hat{e}_1+
v_\perp(\cos\xi\,\hat{e}_1+\sin\xi\,\hat{e}_2)+v_\|\,\hat{n}_\|,\nonumber\\
v_m&=&A\frac{\eta^2(1-q)}{(1+q)}\left[1+B\,\eta\right],\nonumber\\
v_\perp&=&H\frac{\eta^2}{(1+q)}\left[
(1+B_H\,\eta)\,(\alpha_2^\|-q\alpha_1^\|)\right.\nonumber\\
&&\left.+\,H_S\,\frac{(1-q)}{(1+q)^2}\,(\alpha_2^\|+q^2\alpha_1^\|)\right],\nonumber\\
v_\|&=&K\frac{\eta^2}{(1+q)}\Bigg[
(1+B_K\,\eta)
\left|\alpha_2^\perp-q\alpha_1^\perp\right|
\nonumber \\ && \quad \times
\cos(\Theta_\Delta-\Theta_0)\nonumber\\
&&+\,K_S\,\frac{(1-q)}{(1+q)^2}\,\left|\alpha_2^\perp+q^2\alpha_1^\perp\right|
\nonumber \\ && \quad \times
\cos(\Theta_S-\Theta_1)\Bigg],
\end{eqnarray}
where $\eta=q/(1+q)^2$, with $q=m_1/m_2$
the mass ratio of the smaller to larger mass hole,
$\vec{\alpha}_i=\vec{S}_i/m_i^2$, the index $\perp$ and $\|$ refer to
perpendicular and parallel to the orbital angular momentum respectively,
$\hat{e}_1,\hat{e}_2$ are
orthogonal unit vectors in the orbital plane, and $\xi$ measures the
angle between the unequal mass and spin contribution to the recoil
velocity in the orbital plane. 
from newly available runs.
 The angle $\Theta$ is defined as the angle
between the in-plane component of $\vec \Delta = M (\vec S_2/m_2 - \vec
S_1/m_1)$ or $\vec S=\vec S_1+\vec S_2$
and a fiducial direction at merger (see Ref.~\cite{Lousto:2008dn} technique). 
Phases $\Theta_0$ and $\Theta_1$ depend
on the initial separation of the holes for quasicircular orbits
(astrophysically realistic evolutions of comparable masses black holes
lead to nearly zero eccentricity mergers).

The empirical formula (\ref{eq:Pempirical}) above was obtained by
assuming the post-Newtonian dependence on the spin and mass ratio of
instantaneous radiated linear momenta \cite{Kidder:1995zr} where the
coefficients are to be fitted by full numerical simulations.
Second order corrections in the spin have been obtained recently
\cite{Racine:2008kj} and could be added to the empirical formula.
Here, in this paper, we will consider instead the particular family of
configurations that lead to the maximum
recoil~\cite{Campanelli:2007ew, Campanelli:2007cga, Gonzalez:2007hi,
Dain:2008ck}, where $q=1$ and the two spins are in the orbital plane,
equal in magnitude, and opposite in direction.  These configurations
are $\pi-$symmetric, i.e. rotating the system by 180 degrees
around the symmetry axis lead to the same configuration.  This implies
in particular, that only odd powers of the spin and the $\cos\Theta$
are involved. We will then perform a series of simulations that vary
both the magnitude of the (intrinsic) spin in the range
$\alpha=0.2-0.92$ and the initial angle of the individual black-hole
spin and orbital linear momentum.

For a first exploration of the extended spin dependence we consider
cubic and possible fifth-order corrections~\cite{Boyle:2007ru}
 to the empirical formula (\ref{eq:Pempirical}) of the form
\begin{eqnarray}
  v_\| &=& \left(V_{1,1} \alpha + V_{1,3} \alpha^3\right) 
       \cos(\Theta_\Delta-\Theta_0) \nonumber \\
      &+&
        \left(V_{3,1} \alpha + V_{3,3} \alpha^3 + V_{3,5} \alpha^4 \right)
         \cos(3 \Theta_\Delta-3 \Theta_3),
  \label{eq:emp}
\end{eqnarray}
where $V_{1,1} = 2 K(1+\eta B_K) \frac{\eta^2}{(1+q)}$,
and the remaining terms are higher-order correction to
Eq.~(\ref{eq:Pempirical}).

\section{Techniques}
\label{sec:techniques}
To compute the numerical initial data, we use the puncture
approach~\cite{Brandt97b} along with the {\sc
TwoPunctures}~\cite{Ansorg:2004ds} thorn.  In this approach the
3-metric on the initial slice has the form $\gamma_{a b} = (\psi_{BL}
+ u)^4 \delta_{a b}$, where $\psi_{BL}$ is the Brill-Lindquist
conformal factor, $\delta_{ab}$ is the Euclidean metric, and $u$ is
(at least) $C^2$ on the punctures.  The Brill-Lindquist conformal
factor is given by $ \psi_{BL} = 1 + \sum_{i=1}^n m_{i}^p / (2 |\vec r
- \vec r_i|), $ where $n$ is the total number of `punctures',
$m_{i}^p$ is the mass parameter of puncture $i$ ($m_{i}^p$ is {\em
not} the horizon mass associated with puncture $i$), and $\vec r_i$ is
the coordinate location of puncture $i$. For the initial (conformal) extrinsic
curvature we take the analytic form $\hat{K}_{ij}^{BY}$ given by 
Bowen and York\cite{Bowen80}.  We evolve these
black-hole-binary data-sets using the {\sc
LazEv}~\cite{Zlochower:2005bj} implementation of the moving puncture
formalism~\cite{Campanelli:2005dd,Baker:2005vv} with the conformal
factor $W=\sqrt{\chi}=\exp(-2\phi)$ suggested by~\cite{Marronetti:2007wz}
as a dynamical variable.
For the runs presented here
we use centered, eighth-order finite differencing in
space~\cite{Lousto:2007rj} and an RK4 time integrator (note that we do
not upwind the advection terms).

We use the Carpet~\cite{Schnetter-etal-03b} mesh refinement driver to
provide a `moving boxes' style mesh refinement. In this approach
refined grids of fixed size are arranged about the coordinate centers
of both holes.  The Carpet code then moves these fine grids about the
computational domain by following the trajectories of the two black
holes.

We use {\sc AHFinderDirect}~\cite{Thornburg2003:AH-finding} to locate
apparent horizons.  We measure the magnitude of the horizon spin using
the Isolated Horizon algorithm detailed in~\cite{Dreyer02a}. This
algorithm is based on finding an approximate rotational Killing vector
(i.e.\ an approximate rotational symmetry) on the horizon $\varphi^a$. Given
this approximate Killing vector $\varphi^a$, the spin magnitude is
\begin{equation}
 \label{isolatedspin} S_{[\varphi]} =
 \frac{1}{8\pi}\int_{AH}(\varphi^aR^bK_{ab})d^2V,
\end{equation}
where $K_{ab}$ is the extrinsic curvature of the 3D-slice, $d^2V$ is
the natural volume element intrinsic to the horizon, and $R^a$ is the
outward pointing unit vector normal to the horizon on the 3D-slice.
We measure the direction of the spin by finding the coordinate line
joining the poles of this Killing vector field using the technique
introduced in~\cite{Campanelli:2006fy}.  Our algorithm for finding the
poles of the Killing vector field has an accuracy of $\sim 2^\circ$
(see~\cite{Campanelli:2006fy} for details). Note that once we have the
horizon spin, we can calculate the horizon mass via the Christodoulou
formula
\begin{equation}
{m^H} = \sqrt{m_{\rm irr}^2 +
 S^2/(4 m_{\rm irr}^2)},
\end{equation}
where $m_{\rm irr} = \sqrt{A/(16 \pi)}$ and $A$ is the surface area of
the horizon.
We measure radiated energy, linear momentum, and angular momentum, in
terms of $\psi_4$, using the formulae provided in
Refs.~\cite{Campanelli99,Lousto:2007mh}. However, rather than using
the full $\psi_4$, we decompose it into $\ell$ and $m$ modes and solve
for the radiated linear momentum, dropping terms with $\ell \geq 5$.
The formulae in Refs.~\cite{Campanelli99,Lousto:2007mh} are valid at
$r=\infty$.
We obtain highly accurate values for these quantities by
solving for them on spheres of finite radius (typically $r/M=50, 60,
\cdots, 100$), fitting the results to a polynomial dependence in
$l=1/r$, and extrapolating to
$l=0$~\cite{Baker:2005vv,Campanelli:2006gf,Hannam:2007ik,Boyle:2007ft}. Each quantity $Q$ has the radial
dependence $Q=Q_0 + l Q_1 + {\cal O}(l^2)$, where $Q_0$ is the
asymptotic value (the ${\cal O}(l)$ error arises from the ${\cal
O}(l)$ error in $r\, \psi_4$). We perform both linear and quadratic
fits of $Q$ versus $l$, and take $Q_0$ from the quadratic fit as the
final value with the differences between the linear and extrapolated
$Q_0$ as a measure of the error in the extrapolations.

We obtain accurate, convergent waveforms and horizon parameters by
evolving this system in conjunction with a modified 1+log lapse and a
modified Gamma-driver shift
condition~\cite{Alcubierre02a,Campanelli:2005dd}, and an initial lapse
$\alpha(t=0) = 2/(1+\psi_{BL}^{4})$.  The lapse and shift are evolved
with
\begin{subequations}
\label{eq:gauge}
  \begin{eqnarray}
(\partial_t - \beta^i \partial_i) \alpha &=& - 2 \alpha K,\\
 \partial_t \beta^a &=& (3/4) \tilde \Gamma^a - \eta(x^a,t) \beta^a,
 \label{eq:Bdot}
 \end{eqnarray}
 \end{subequations}
where different functional dependences for $\eta(x^a,t)$ have been
proposed in
\cite{Alcubierre:2004bm, Zlochower:2005bj, Mueller:2009jx, Mueller:2010bu, Schnetter:2010cz,Alic:2010wu}.
For the low-spin simulations we used a constant $\eta=2$, while for
the $\alpha=0.92$ simulation we used 
a modification of the form proposed
in~\cite{Lousto:2010qx},
\begin{equation}
  \eta(x^a,t) =  R_0 \frac{\sqrt{\tilde
\gamma^{ij}\partial_i W \partial_j W }}{ \left(1 - W^a\right)^b},
\end{equation}
where we chose $R_0=1.31$~\cite{Mueller:2009jx}.
In practice we used
$a=2$ and $b=2$, which reduces $\eta$ by a factor of $4$ at infinity
when compared to the gauge proposed
by~\cite{Mueller:2009jx}, improving its stability at larger radii. Other values
of $(a,b)$ lead to an increase of the numerical noise. 
Note that this gauge was originally proposed and used for the non-spinning, intermediate-mass-ratio binaries. Here we find that the
gauge is well adapted for the highly-spinning equal mass case, where,
after the initial burst of radiation passes,
the measured spin is found to never drop below $\alpha=0.905$.
Due to the differences in the spurious initial radiation content, 
as well as spin-orbit effects on the
total mass, $\alpha$ near merger  varied from between $0.90$ to $0.93$
for the different A09Tyyy configurations 
(See tables \ref{tab:ID} and \ref{tab:rad} below).

\subsection{Initial Data}
\label{sec:ID}

We used 3PN parameters for quasicircular orbits with BH spins 
(equal in magnitude and opposite in direction) aligned
with the linear momentum of each BH (i.e. in-plane spins) to obtain the
momenta and spin parameters for the Bowen-York extrinsic curvature.
We then chose puncture mass parameters such that the total ADM mass was
1M. We then rotated the spins by $30^\circ$, $90^\circ$, $130^\circ$,
$210^\circ$ and $315^\circ$, to obtain a total of 6 configurations
for each value of the intrinsic spin $\alpha$. We label the configuration
AxxTyyy where xx corresponds to the spin of each BH and yyy is the
initial rotation of the spin directions. We summarize the
initial data in Table~\ref{tab:ID}.

\begin{table}
\caption {Initial data parameters for the non-rotated configurations.
The initial puncture positions are $\pm (x,0,0)$, momenta are $\pm(0,p,0)$,
and spin $\pm(0,S,0)$. The remaining configurations are obtained by
rotating the spins, keeping all other parameters the same.}
\label{tab:ID}
\begin{ruledtabular}
\begin{tabular}{lcccc}
Config & $x$ & $p$ & $S$ & $m_p$\\
\hline
A02T000 & 3.878113 & 0.117404 & 0.051314 & 0.479782 \\
A04T000 & 3.879566 & 0.117405 & 0.102627 & 0.454076\\
A06T000 & 3.881979 & 0.117407 & 0.153936 & 0.403550\\
A08T000 & 3.885342 & 0.117409 & 0.205241 & 0.301026\\
A09T000 & 3.887375 & 0.117411 & 0.230891 & 0.172120
\end{tabular}
\end{ruledtabular}
\end{table}

This family of configurations has larger initial separations than the
configurations in our original studies in \cite{Campanelli:2007cga}.
In these configurations, the BHs orbit $\sim 3.5$ times prior to
merger, which allows for most of the eccentricity to be radiated away
before the plunge phase (where most of the recoil velocity is
generated). This provides for an accurate description of the plausible
astrophysical maximal recoil scenario.

\section{Results and Analysis}
\label{sec:results}

In order to analyze our results for different initial orientations of the
spin that span the $\Theta-$dependence, we use the techniques detailed
 in~\cite{Lousto:2008dn}. For each $\alpha$ we fit the results of
the recoil as a function of angle to form 
$V_{\rm recoil} = V_1 \cos(\theta - \theta_1) +
      V_3 \cos[3(\theta - \theta_3)],$ where $\theta$ is defined to be the
angle of the spin direction (of the first BH) near merger (at a
fiducial radial separation of $r=1.2$) and
the spin direction of the corresponding AxxT000 configuration
(we cannot simply use the initial spin direction differences
 because spin-orbit
effects for larger spins make this approximation inaccurate).
The radiated energy and recoil from each simulation is given in
Table~\ref{tab:rad}.

\begin{widetext} 

\begin{table}
\caption{The radiated energy, recoil velocity, and angle between
the spins for the AxxTyyy configuration at merger and the
corresponding AxxT000 configuration, $\Delta\Theta$. Note the substantial
rotations apparent in the A09Tyyy configurations due to
spin orbit interactions.}
\label{tab:rad}
\begin{ruledtabular}
\begin{tabular}{lcccc}
Config & $\delta E$ & $V_{\rm recoil}$ & $\Delta\Theta$ & $\delta J_z$ \\
\hline
A02T000 & $ 0.03583 \pm 0.00020 $ & $ 551.95 \pm 0.83 $ & 0  &
$0.2727 \pm 0.0032 $\\
A02T030 & $ 0.03575 \pm 0.00020 $ & $ 225.49 \pm 0.86 $ & 30.7 &
$0.2724\pm0.0031$ \\
A02T090 & $ 0.03562 \pm 0.00019 $ & $ -482.10 \pm 0.13 $ & 88.6 &
$0.2721\pm0.0031$ \\
A02T130 & $ 0.03574 \pm 0.00019 $ & $ -721.03 \pm 0.34 $ & 127.3 &
$0.2727\pm0.0031$ \\
A02T210 & $ 0.03573 \pm 0.00020 $ & $ -234.14 \pm 0.76 $ & 210.0 &
$0.2723\pm0.0032$ \\
A02T315 & $ 0.03577 \pm 0.00019 $ & $ 730.10 \pm 0.32 $ & 312.4 &
$0.2729\pm0.0030$ \\
\hline

A04T000 & $ 0.03665 \pm 0.00022 $ & $ 1200.79 \pm 2.21 $ & 0 &
$0.2768\pm0.0031$ \\
A04T030 & $ 0.03625 \pm 0.00022 $ & $ 529.09 \pm 2.12 $ & 33.7 &
$0.2747\pm0.0031$ \\
A04T090 & $ 0.03574 \pm 0.00020 $ & $ -764.08 \pm 0.39 $ & 85.0 &
$0.2740\pm0.0028$ \\
A04T130 & $ 0.03620 \pm 0.00021 $ & $ -1390.22 \pm 0.89 $ & 126.0 &
$0.2759\pm0.0029$ \\
A04T210 & $ 0.03633 \pm 0.00021 $ & $ -637.049 \pm 1.97 $ & 209.2 &
$0.2755\pm0.0030$ \\
A04T315 & $ 0.03628 \pm 0.00021 $ & $ 1424.17 \pm 1.02 $ & 311.2 &
$0.2762\pm0.0029$ \\
\hline

A06T000 & $ 0.03803 \pm 0.00027 $ & $ 2001.06 \pm 3.5 $ & 0 &
$0.2834\pm0.0032$ \\
A06T030 & $ 0.03740 \pm 0.00026 $ & $ 1087.24 \pm 4.8 $ & 36.2&
$0.2798\pm0.0032$ \\
A06T090 & $ 0.03595 \pm 0.00025 $ & $ -870.93 \pm 2.4 $ & 87.3&
$0.2758\pm0.0026$ \\
A06T130 & $ 0.03669 \pm 0.00026 $ & $ -1944.04 \pm 1.5 $ & 127.7&
$0.2791\pm0.0029$ \\
A06T210 & $ 0.03751 \pm 0.00026 $ & $ -1212.62 \pm 4.4 $ & 212.4&
$0.2807\pm0.0031$ \\
A06T315 & $ 0.03679 \pm 0.00026 $ & $ 1984.74 \pm 1.3 $ & 310.3&
$0.2797\pm0.0029$ \\
\hline

A08T000 & $ 0.03996 \pm 0.00039 $ & $ 2651.75 \pm 7.54 $ & 0 &
$0.2912\pm0.0027$ \\
A08T030 & $ 0.03941 \pm 0.00037 $ & $ 1917.03 \pm 8.29 $ & 24.3 &
$0.2888\pm0.0027$ \\
A08T090 & $ 0.03677 \pm 0.00034 $ & $ -445.83 \pm 1.10 $ & 70.9&
$0.2791\pm0.0028$ \\
A08T130 & $ 0.03733 \pm 0.00038 $ & $ -2412.2 \pm 4.03 $ & 119.5&
$0.2823\pm0.0027$ \\
A08T210 & $ 0.03941 \pm 0.00036 $ & $ -1919.86 \pm 8.26 $ & 204.3&
$0.2887\pm0.0027$ \\
A08T315 & $ 0.03771 \pm 0.00038 $ & $ 2568.87 \pm 4.49 $ & 306.3&
$0.2838\pm0.0027$ \\
\hline

A09T000 & $ 0.04026 \pm 0.00057 $ & $ 89.74 \pm 1.00 $ & 0 &
$0.3063\pm0.0042$ \\
A09T030 & $ 0.04143 \pm 0.00055 $ & $ 3240.96 \pm 17.34 $ & 101.5&
$0.3011\pm0.0051$ \\
A09T090 & $ 0.04062 \pm 0.00057 $ & $ 1859.42 \pm 15.47 $ & 147.4&
$0.2951\pm0.0070$ \\
A09T130 & $ 0.03784 \pm 0.00054 $ & $ -759.16 \pm 0.85 $ & 190.6&
$0.2846\pm0.0058$ \\
A09T210 & $ 0.04144 \pm 0.00055 $ & $ -3239.25 \pm 17.24 $ & 281.7&
$0.3012\pm0.0050$ \\
A09T315 & $ 0.03917 \pm 0.00056 $ & $ -205.59 \pm 2.92 $ & 355.3&
$0.2919\pm0.0067$ \\
\end{tabular}
\end{ruledtabular}
\end{table}

\end{widetext} 
We then fit $V_1$ and $V_3$ to the functional forms $V_1 = V_{1, 1}
 \alpha + V_{1, 3} \alpha^3$ and $V_3 = V_{3, 1} \alpha + V_{3, 3} \alpha^3$.
A summary of the fits is given in Table~\ref{tab:fits}. Note that
$V_{1,1}$ is related to the parameter $K$ in our empirical formula~(\ref{eq:emp}) by $K=16 V_{1,1}$. Here we find $K=58912\pm43$, where the error is
obtained from the fit and likely underestimated the true error in this
quantity. Previously we found $K=(6.0\pm0.1)\times10^4$, which agrees
reasonably well with the new value
\cite{Campanelli:2007cga, Lousto:2008dn}. We also include fits
where the linear term in $V_3$ and the cubic term in $V_1$ are
set to zero, as well as a fit of $V_3$ to $V_{3,3}\alpha^3 + V_{3,5}
\alpha^5$. We note that a cubic term in $V_1$ is expected
since $\cos^3 \theta = 3/4 \cos \theta + 1/3 \cos 3 \theta$, and
hence cubic corrections of the for $\alpha^3 \cos^3\theta$ will
contribute to the $\cos\theta$ dependence. On the other hand,
a linear dependence in $\cos 3\theta$ is not expected.

The form of the fitting above was first proposed in \cite{Boyle:2007ru} as a
generic expansion, where it was applied to data sets with constant
$\alpha$. Here we compare results from five different values of the 
intrinsic
spin in the range $\alpha=0.2-0.92$, to obtain an accurate model of
the $\alpha$ dependence.

\begin{figure}
  \caption{Fit of the recoil versus angle for the $\alpha=0.2$
    configurations.}
  \label{fig:fit_a2_ang}
   \includegraphics[width=3in]{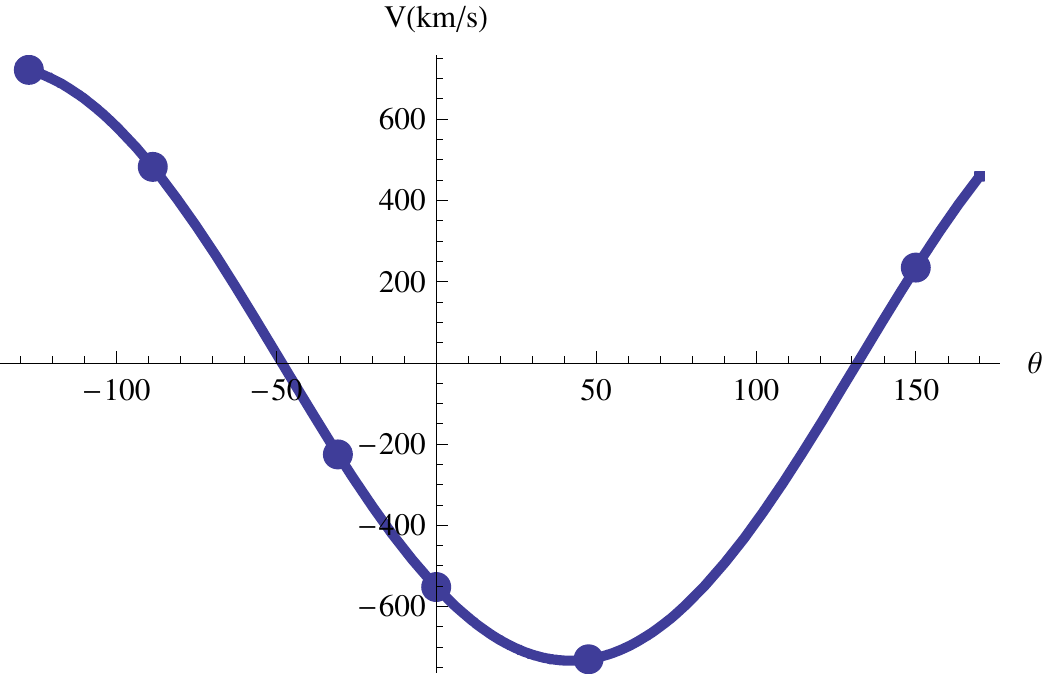}
\end{figure}

\begin{figure}
  \caption{Fit of the recoil versus angle for the $\alpha=0.4$
    configurations.}
  \label{fig:fit_a4_ang}
   \includegraphics[width=3in]{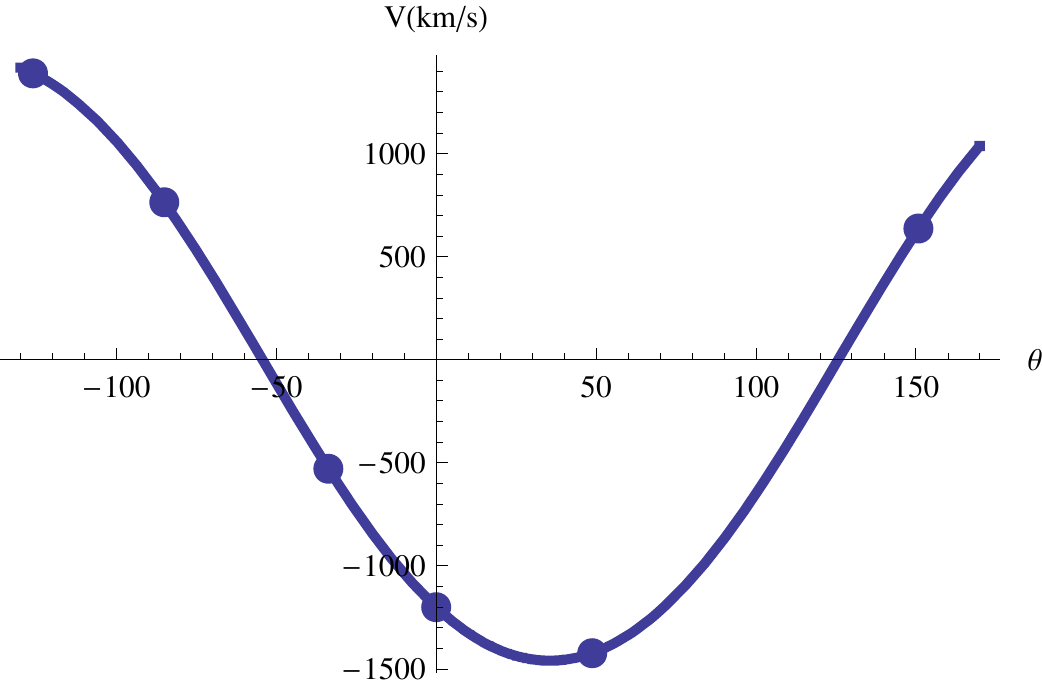}
\end{figure}

\begin{figure}
  \caption{Fit of the recoil versus angle for the $\alpha=0.6$
    configurations.}
  \label{fig:fit_a6_ang}
   \includegraphics[width=3in]{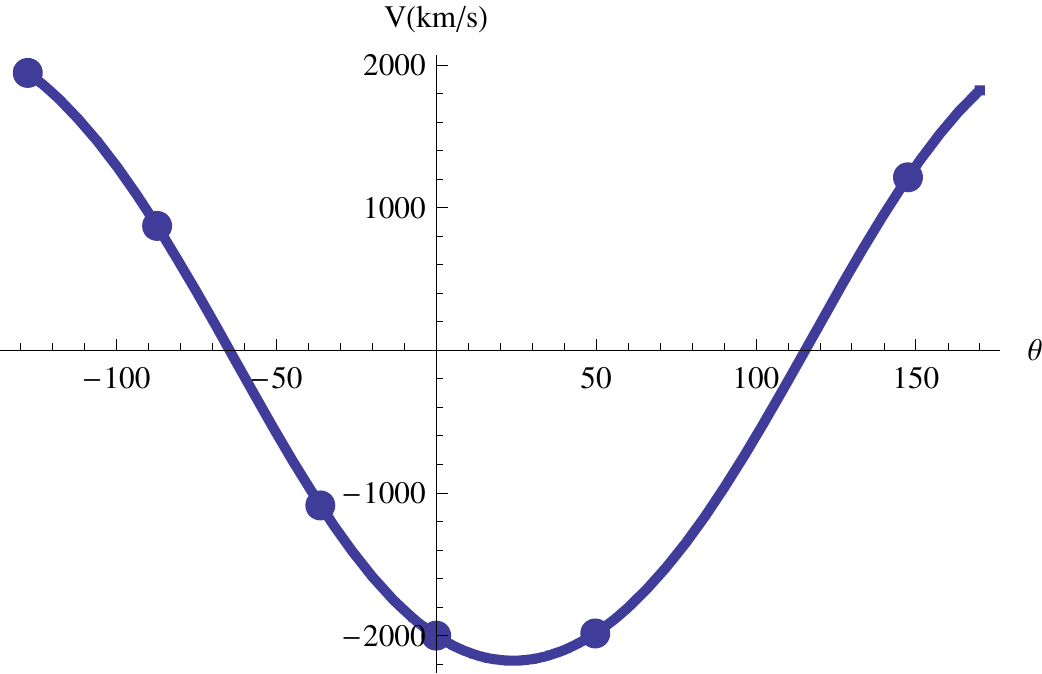}
\end{figure}

\begin{figure}
  \caption{Fit of the recoil versus angle for the $\alpha=0.8$
    configurations.}
  \label{fig:fit_a8_ang}
   \includegraphics[width=3in]{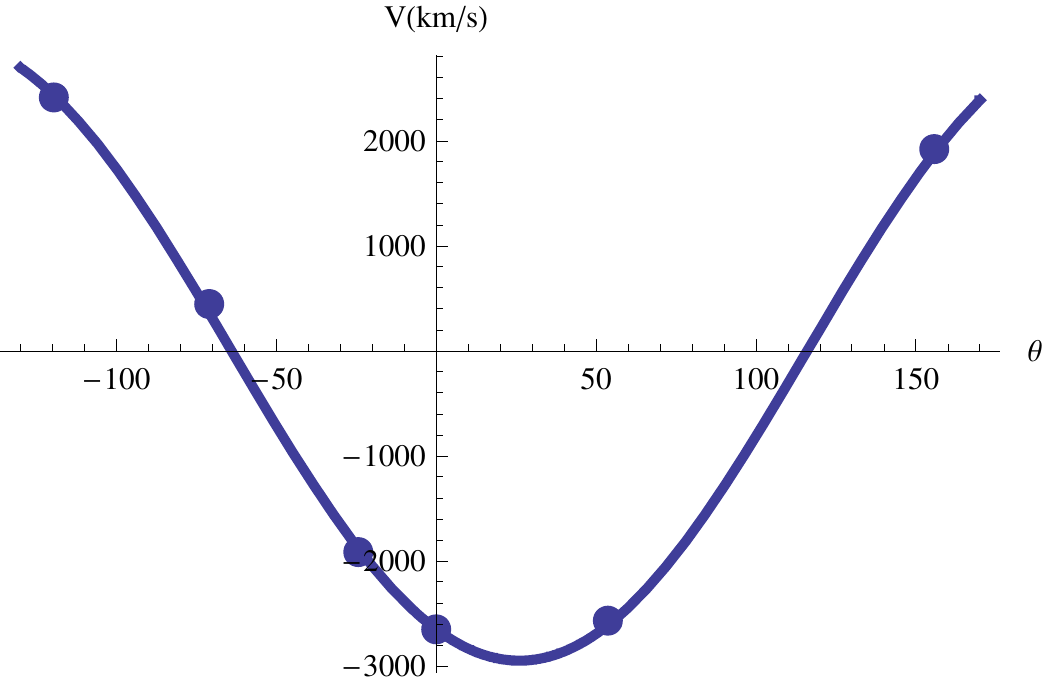}
\end{figure}

\begin{figure}
  \caption{Fit of the recoil versus angle for the $\alpha=0.92$
    configurations.}
  \label{fig:fit_a9_ang}
   \includegraphics[width=3in]{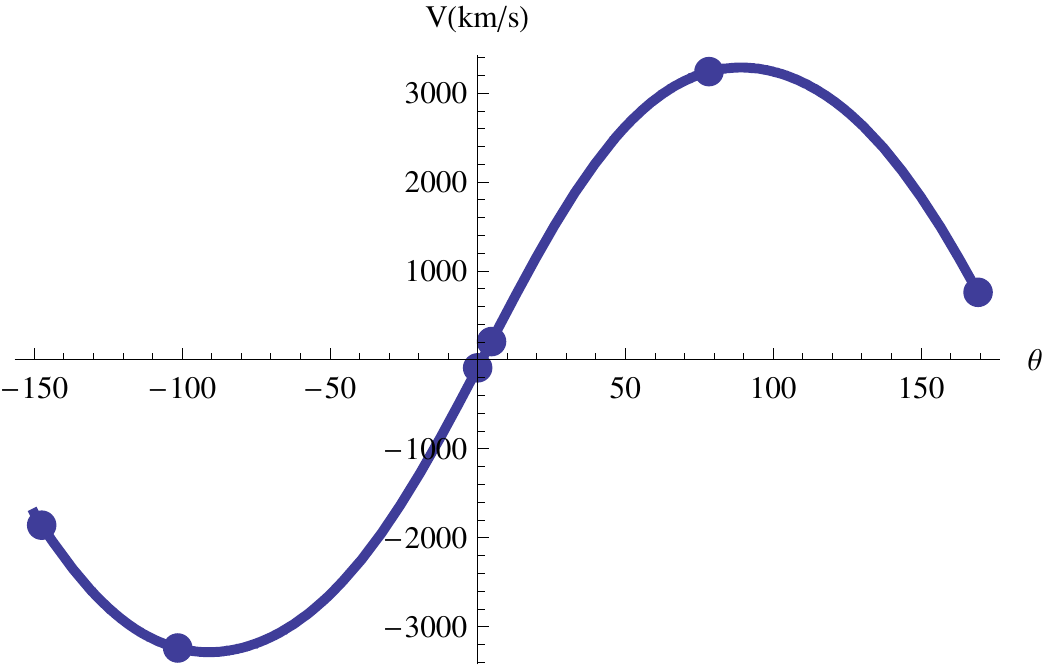}
\end{figure}

In Figs.~\ref{fig:fit_a2_ang}-\ref{fig:fit_a9_ang}
 we show the angular fits for each set of
configurations. Note that the spin-orbit coupling effects are
strongest for the A09Tyyy configurations, as is apparent by the
relative translation of two configurations towards the same final
angle.

\begin{table}
\caption {Fits of the recoil to the functional form
$v_{\rm kick} = V_1 \cos(\theta - \theta_1) +
      V_3 \cos[3(\theta - \theta_3)].$ Note that angles are
measured in degrees. The reported errors come from the nonlinear
least-squares fit of the data and are underestimates of the
actual errors. Note the average value of $\alpha$ for the
A09Tyyy configuration near merger was $\alpha\sim0.92$, which
was the value used to obtain the fits to
$V_{i,j}$.}
\label{tab:fits}
\begin{ruledtabular}
\begin{tabular}{lcc}
$\alpha$  & $V_1 $ & $\theta_1$ \\
\hline
0    & 0       & **** \\
0.2  & $737.70 \pm 0.12$ & $221.8002\pm0.0010$ \\
0.4  & $1472.59\pm0.06$ & $215.6909\pm0.0011$ \\
0.6  & $2204.98\pm0.56$ & $205.117\pm0.015$ \\
0.8  & $2935.93\pm0.65$ & $206.658\pm0.013$ \\
0.9  & $3376.3\pm7.5$ & $91.02\pm0.11$ \\
\hline
\hline
$\alpha$ & $V_3$ & $\theta_3$ \\
\hline
0     0        & **** \\
0.2   & $4.23\pm 0.12$ & $279.62\pm 0.65$\\
0.4   & $12.0838\pm0.024$ & $37.790\pm0.049$\\
0.6  & $31.63\pm0.55$ & $152.72\pm0.38$\\
0.8  &  $69.21\pm0.74$ & $38.01\pm0.22$ \\
0.9  &  $95.5\pm2.4$ & $36.7\pm1.5$\\
\hline
$V_{1,1}$ & $3681.77\pm2.66$ \\
$V_{1,3}$ & $-15.46\pm3.97$ \\
$V_{3,1}$ & $15.65\pm3.01$ \\
$V_{3,3}$  & $105.90\pm4.50$

\end{tabular}
\end{ruledtabular}
\end{table}

In Table~\ref{tab:fits} we provide the fitting constants $V_{i,j}$
assuming the spin of the A09Tyyy was $0.92$. in actuality, the spin
varied between configurations. In Table~\ref{tab:varya} we provide
fitting parameters for $V_{i,j}$ if we take the value of $\alpha$ 
for these configurations to be $\alpha=0.9$ (the expected value when
neglecting effects due to the initial radiation content),
$\alpha=0.91$, and $\alpha=0.92$ (which approximates the average value
of $\alpha$ over all configurations). We find that setting
$\alpha=0.92$ gives the best fit for the dominant $V_{1,1}$ term.
However, we note that these fits do indicate that the nonleading
$V_{1,3}$ term and $V_{3,1}$ term may be zero. We therefore also
provide fits assuming these two terms vanish. Fits to $V_1$ strongly
prefer $\alpha=0.92$ over the smaller values. We note that
the sign of $V_{1,3}$ changes if we assume smaller values of $\alpha$
for the A09Tyyy configurations.

\begin{table}
\caption {Fits $V_1$ and $V_3$ to the form
  $V_1=V_{1,1} \alpha + V_{1,3} \alpha^3$ and
  $V_3 = V_{3,1} \alpha + V_{3,3} \alpha^3$,
as well as $V_3 = V_{3,3} \alpha^3 + V_{3,5} \alpha^5$. For the 
A09Tyyy configurations we take $\alpha=0.9$, $0.91$, and $0.92$, which
accounts for the expected value of $\alpha$ for these configuration,
the actual average value observed, and a spin between these two
values, as
explained in the text. $\delta^2$ is the average of the square of the
error in the fit. Note that fits to the dominant $V_1$ term strongly
prefer $\alpha=0.92$ over smaller values, while fits to the subleading
$V_3$ prefer $\alpha=0.9$.}
\label{tab:varya}
\begin{ruledtabular}
\begin{tabular}{lccc}
$\alpha$ (A09Tyyy) & $V_{1,1}$ & $V_{1,3}$ & $\delta^2$ \\
\hline
0.92 & $3681.77\pm2.66$ & $-15.46\pm3.966$ & 1.21 \\
0.91 & $3658.21\pm 20.74$ & $49.16, 31.47$ & 71.13 \\
0.90 & $3634.85\pm41.09$  & $115.31, 63.40$ & 270.25 \\
\hline
\hline
$\alpha$ (A09Tyyy) & $V_{3,1}$ & $V_{3,3}$ & $\delta^2$ \\
\hline
0.92 & $15.65\pm 3.01$ & $105.90\pm 4.50$ & 1.55 \\
0.91 & $13.68\pm 1.82$ & $111.15\pm 2.77$ & 0.55 \\
0.90 & $11.75\pm 1.14$ & $116.45\pm 1.77$ & 0.21 \\
\hline\hline
$\alpha$ (A09Tyyy) & $V_{1,1}$ & &  $\delta^2$\\
\hline
0.92  & $3672.08\pm 1.84$ & 0& 5.80 \\
0.91  & $3688.56\pm 8.23$ & 0& 114.53 \\
0.90  & $3704.98\pm 17.17$ &0 & 493.78 \\
\hline\hline
$\alpha$ (A09Tyyy) & & $V_{3,3}$ &  $\delta^2$\\
\hline
0.92  & 0& $127.74\pm 3.96$ & 12.02\\
0.91  & 0& $130.60\pm 3.36$ & 8.29\\
0.90  & 0& $133.45\pm 2.90$ & 5.73 \\
\hline
$\alpha$ (A09Tyyy) & $V_{3,3}$ & $V_{3,5} $&  $\delta^2$\\
\hline
0.92  & $172.55\pm10.20$ &  $-58.98\pm13.21$ & $2.01$ \\
0.91  & $167.45\pm10.87$ & $-49.54\pm14.38$ & $2.09$ \\
0.90  & $161.80\pm12.15$ & $-38.88\pm16.43$ & $2.389 $

\end{tabular}
\end{ruledtabular}
\end{table}

\begin{figure}
  \caption{A comparison of fits of  $V_3$ to
    $V_3 = \alpha V_{3,1} + \alpha^3 V_{3,3}$ (solid),
    $V_3 = \alpha^3 V_{3,3} + \alpha^5 V_{3,5}$ (dotted),
    $V_3 = \alpha^3 V_{3,3}$ (dot-dashed). The first fit
    is the best.
    In all cases the spins for the A09Tyyy configurations
     were assumed to be $\alpha=0.92$}
  \label{fig:v3_fit_comp}
  \includegraphics[width=3in]{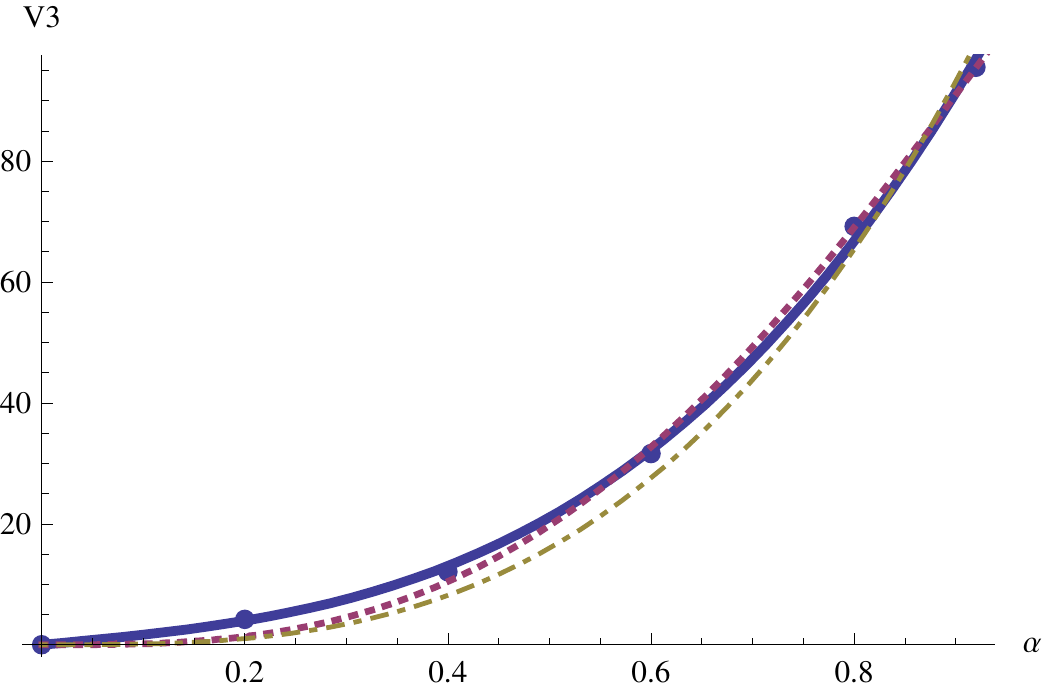}
\end{figure}

\begin{figure}
  \caption{A comparison of fits of  $V_1$ to
    $V_1 = \alpha V_{1,1} + \alpha^3 V_{1,3}$ (solid),
    $V_3 = \alpha V_{1,1}$ (dotted).
    There is no significant differences between the fits.
    In all cases the spins for the A09Tyyy configurations
     were assumed to be $\alpha=0.92$}
  \label{fig:v1_fit_comp}
  \includegraphics[width=3in]{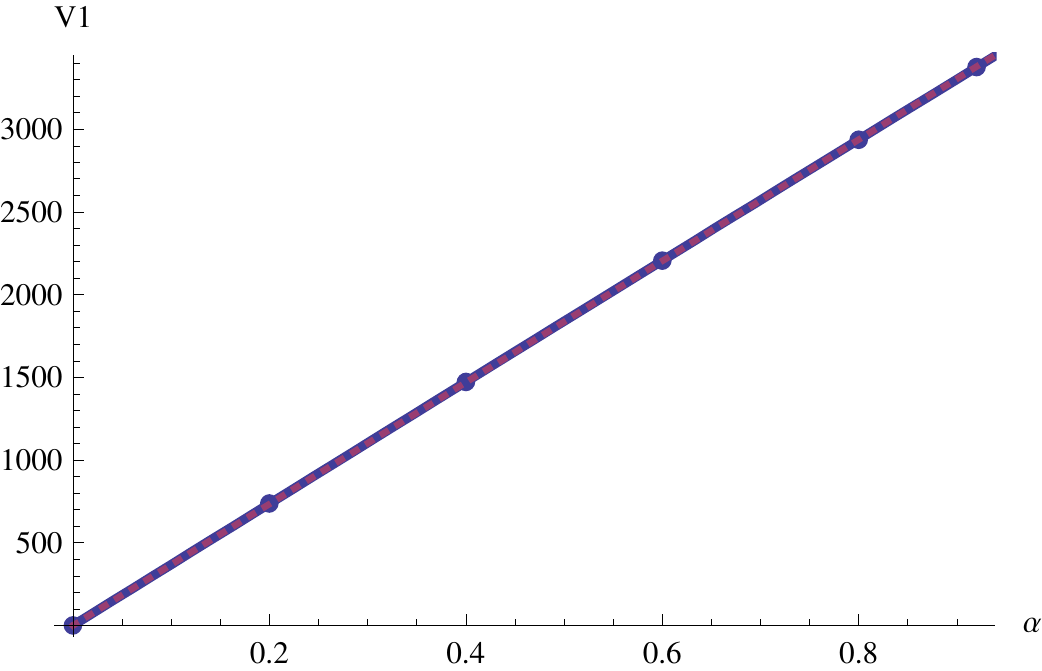}
\end{figure}

While arguments based on post-Newtonian scaling do not seem
to indicate the presence of an $\alpha \cos 3\Theta$ term, our results
indicate that this term is present. This may indicate an
error in $V_3$ for $\alpha=0.2$. If we exclude this data point,
then we can fit to reasonably well to either $V_{3,1} \alpha + V_{3,3}
\alpha^3$ or $V_{3,3} \alpha^3 + V_{3,5} \alpha^5$. Further
exploration in the small $\alpha$ regime is required.
 In Figs~\ref{fig:v3_fit_comp} we
compare fits of $V_3$ and find that the best fit is
to $V_3 = \alpha V_{3,1} + \alpha^3 V_{3,3}$. On the other hand,
as seen in Fig.~\ref{fig:v1_fit_comp}, there is no 
significant difference apparent in the fits of 
$V_1$ to $V_{1,1} \alpha + V_{1,3} \alpha^3$ and
$V_1 = V_{1,1} \alpha$.

\section{Conclusion}
\label{sec:discussion}

Using the enhanced recoil formula for the ``maximum kick'' configurations,
we predict that the maximum recoil will be $3680\pm130 \KMS$, where
the error in the prediction is due to the possibility of the
higher-order effects producing recoils in the same direction
or opposite direction of the dominant linear contribution.
We also established a model for higher-order dependences on the
spin in the recoil formula. These results are particularly relevant
for the interpretation of observations of emission lines in AGNs 
displaying
displacements between narrow and wide emission lines  of the order of
thousands of kilometers per second. In particular in Ref.~\cite{Civano:2010es}
a 1200 km/s offset velocity was measured (CXOCJ100043.1+020637).
A 2650 km/s recoiling supermassive black hole could explain the 
observations (SDSS J092712+294344)
of Ref.~\cite{Komossa:2008qd}. 
While in Ref.~\cite{Shields:2009jf} (SDSS J105041+345631)
and in Ref.~\cite{Boroson:2009va} (SDSS J153636+044127)
there is speculation that 3500 km/s recoiling black holes
are responsible for these features in the spectra.
While none of those cases effectively surpasses the maximum recoil
velocity determined here, they came close enough for the probability
of actually observing this event to be very low \cite{Lousto:2009ka} 
thus leading
to the question about what are the astrophysical mechanisms responsible of
generating such large differential velocities 
\cite{Vivek:2009mm,Lauer:2009us}.

\acknowledgments
We gratefully acknowledge the NSF for financial support from Grants
No. PHY-0722315, No. PHY-0653303, No. PHY-0714388, No. PHY-0722703,
No. DMS-0820923, No. PHY-0929114, No. PHY-0969855, No. PHY-0903782,
No. CDI-1028087; and NASA for financial support from NASA Grants
No. 07-ATFP07-0158 and No. HST-AR-11763.  Computational resources were
provided by the Ranger cluster at TACC (Teragrid allocation TG-PHY060027N)
and by NewHorizons at RIT.

\bibliographystyle{apsrev}
\bibliography{../../../Bibtex/references}

\end{document}